# Integrated Programmable Controlled Phase Gate Design for Quantum Information Processing

Yalın Başay[1, *], Serdar Kocaman[1]

[1]*Middle East Technical University*
*basay.yalin@metu.edu.tr*

## Abstract

In this paper, we proposed a design of integrated programmable controlled-phase (CPHASE) gate to be used in quantum information processing applications. This gate is capable of introducing arbitrary phase difference to target qubit in the case of control qubit being in the state of |1>. Because the desired phase difference is possible to be utilized after fabrication unlike the conventional controlled-phase gates which provide hard-coded phase shift, such an integrated gate is expected to pave the way for more versatile operations of current integrated optical circuits as well as possible new applications.

## Introduction

Research on quantum information technologies is an up-to-date occupation for scientists and engineers since it shows promise to yield unique applications by taking the advantage of quantum phenomena such as superposition and entanglement. These efforts started to yield outcomes that can't be classically achieved [1], [2] and although these results have limited area of usage, more crucial applications especially in the fields of cryptography and secure communication are expected to be achieved in near future [3]. Beside long-standing efforts of implementing quantum information processing via trapped ions [4], [5] or superconductors [6], [7], linear optical quantum information processing, which is a relatively new field, demonstrates an outstanding performance. Photons are preferable carriers of quantum information having little interaction with environment. On the other hand, they don't directly interact with other photons, thereby it is difficult to implement two qubit gates that are essential to achieve universal quantum computation. However, an interaction-like behavior using ancilla photons makes photonic two qubit gates achievable [8], [9]. This leads to vast amount of research on linear optical quantum information processing that offers advantages as more durability of photons to decoherence and potential to link quantum computation and quantum communication in the same framework. By encoding quantum information to various degrees of freedom of photons such as polarization [10]-[12], orbital angular momentum [13]-[15], time-bin [16]-[18], frequency [19], [20] and path [21]-[23], researchers have been making great progress towards large scale photonic quantum computation as well as applications as quantum key distribution [24], quantum random number generation [25] and quantum simulation [26], [27].

Moreover, quantum advantage in some nonuniversal tasks as boson sampling has been successfully demonstrated by using linear optical elements [2]. On the other hand, to achieve universal quantum computation, a complete set of universal gates should be implemented. The focus of most of the researches is realization of controlled-not (CNOT) gate [28], [29] usage of which with single qubit rotations leads to universal quantum computation, or SWAP gate [30]-[32] which eases implementation, however CPHASE gate is also a prominent candidate to achieve this [33]. An advantage of CPHASE gate is that it is possible to introduce desired phase difference between two quantum states. This phase difference can either be fixed [34] or determined during the operation [35]. The latter case provides more versatile quantum circuits by introducing program qubits aim of which is to supply desired phase difference. Offering miniaturized and more stable structures compared to bulk counterparts, integrated quantum photonics is a promising platform to achieve scalable quantum operations [36]. In literature, there are several works focusing on integration of CNOT gates and circuits based upon them [37], [38], while less study on other two qubit gates as CPHASE gate [39]. In this paper, we demonstrate integrated optical implementation of CPHASE gate with programmable phase shift. We used a unique and practical polarization rotator that was recently proposed by Xu et al. to implement half-wave plates in this scheme [40].

What expected from a tunable CPHASE gate is to introduce an arbitrary phase shift which is determined by program qubit, in the case of both target and control qubits being in the state of |1>. In other words, it adds a desired phase difference between |0> and |1> states of target qubit if control qubit is in the state of |1>. This operation can be achieved probabilistically but heralded by using polarization encoded qubits. In [35], an optical setup for tunable CPHASE operation is designed such a way that success of the operation depends on observation of one photon at each of the target and control outputs, and detection of a photon on a detector. Scheme of such a circuit is shown at Figure 1. Optical

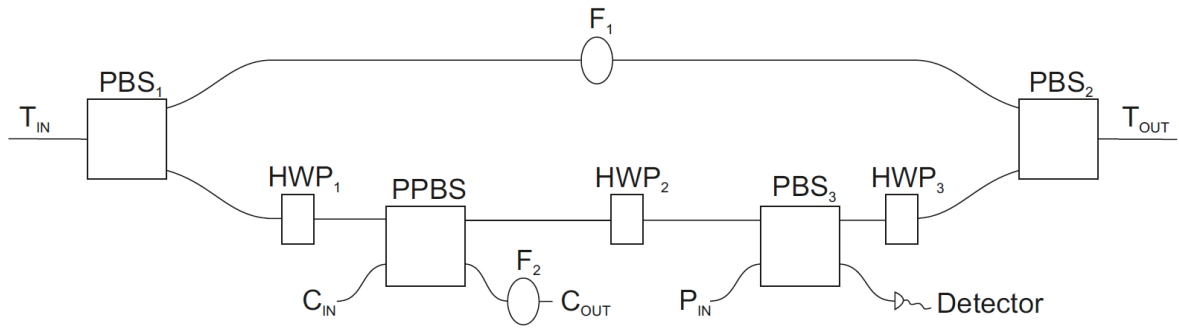

*Figure 1 Scheme of programmable controlled phase gate.*

elements used in this scheme are three polarizing beam splitters ($PBS_1$, $PBS_2$, $PBS_3$) which transmit horizontally polarized photons while reflecting vertically polarized ones, one partially polarizing beam splitter (PPBS) which has unit transmissivity for horizontal polarization while transmissivity of $\frac{1}{\sqrt{3}}$ for vertical polarization, two filters ($F_1$, $F_2$) one of which has amplitude transmissivity of $\frac{1}{2}$ ($F_1$) while the other one only filters horizontal polarization with transmissivity of $\frac{1}{\sqrt{3}}$ ($F_2$), and finally three half-wave plates ($HWP_1$, $HWP_2$, $HWP_3$) transformations performed by which are following:

For $HWP_1$:

$$|1> \rightarrow \frac{1}{2}|0> + \frac{\sqrt{3}}{2}|1>$$

For $HWP_2$ and $HWP_3$:

$$|0> \rightarrow \frac{1}{\sqrt{2}}(|0> + |1>)$$

$$|1> \rightarrow \frac{1}{\sqrt{2}}(|0> - |1>)$$

In this scheme, incoming photon from $T_{IN}$ is transmitted to upper arm if it is horizontally polarized or reflected to lower arm otherwise after input to $PBS_1$. After this separation of states, horizontally polarized state is subjected to $F_1$ which filter some portion of its amplitude to equalize output amplitude with amplitude of states processed at lower arm. In the mean time, vertically polarized state at lower arm reaches $HWP_1$ and transforms to superposition of two polarization states which is a requirement for successful operation since if $C_{IN}$ is in |0> state while other input of PPBS is |1>, both photons will output from $C_{OUT}$ so no successful operation would be possible for this combination. The most critical part of scheme is PPBS where quantum interference takes place if both target and control qubits are in |1> state. By this way, a phase difference of π emerges between |0> and |1> states. As a result of this phase difference, outcome of $HWP_3$ becomes |1>, which would be |0> if there was no phase difference, and the |1> state is detected by detector since it is reflected at $PBS_3$. Because success of the operation depends on detection at detector,

which is satisfied now, |1> state of program qubit that has desired phase enters the lower arm of the circuit and output from T$_{OUT}$ after being subjected to HWP$_3$. The function of HWP$_3$ here is to ensure some portion of |0> state turning into |1> state to be reflected at PBS$_2$ and outcome from T$_{OUT}$. The detailed mathematical steps of operation are explained in [35].

## Design and Results

To construct this scheme as a photonic integrated circuit, all elements mentioned above should be designed appropriately and then combined. Our design tool is finite-difference-time-domain based program MEEP and we performed simulations for 1.55 μm wavelength of photons and silicon-on-insulator (silica is used as an insulator) waveguides, height and width of which are both 350 nm. Directional couplers are used as beamsplitter and in coupling regions, separation distance between two waveguides is set to be 250 nm. Transmissivity and reflectivity of beam splitters can be implemented by changing lengths of the couplers since coupling rates are different for vertical and horizontal polarizations as a result of slight birefringence as shown in Figure 2. Here length of one cycle is 35.80 μm for E$_x$ (horizontal polarization) and 8.32 μm for E$_z$ (vertical polarization). From calculations, for PBS$_1$, PBS$_2$, PBS$_3$ the coupling length turns out to be 70.72 μm and for PPBS it turns out to be 35.90 μm. Similar approach is useful to construct integrated filters function of which is to get rid of some proportion of light. By transferring the desired proportion of light to a waveguide which removes it from the circuit this purpose can be achieved. For F$_1$, only horizontal polarization can be considered since no vertical polarization passes through the filter. The length of coupling region after which energy of horizontal polarization drops to one fourth of initial value is 12.00 μm. Lastly for $F_2$, one-third of the energy of horizontal polarization and all of the vertical polarization should stay in initial waveguide. By performing similar calculations, it turns out that 83.20 μm is the length for such operation.

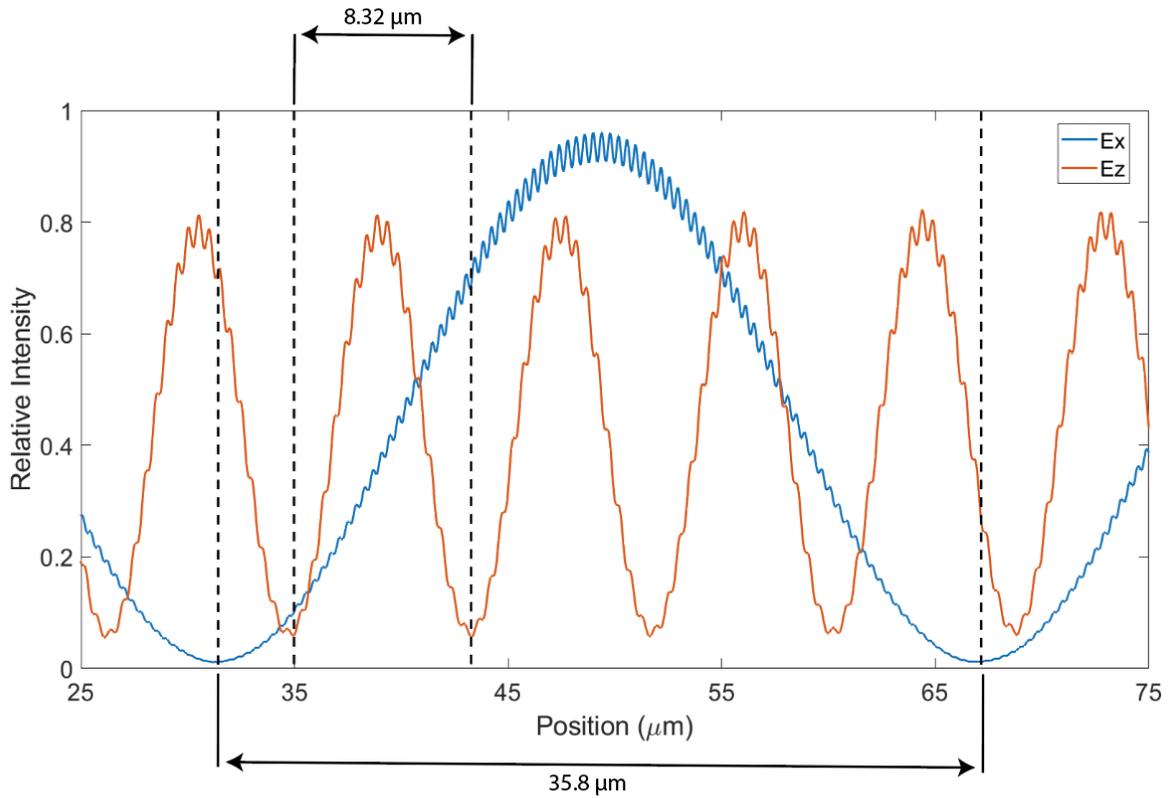

*Figure 2 Relative intensities of horizontal ($E_x$) and vertical ($E_z$) polarization at coupling region where two waveguides are placed with 250 nm distance in between.*

As half-wave plates, we use the notched ring resonator recently proposed at [40]. To ensure all of the light transferring to structure, geometry shown in Figure 3-a is used. Note

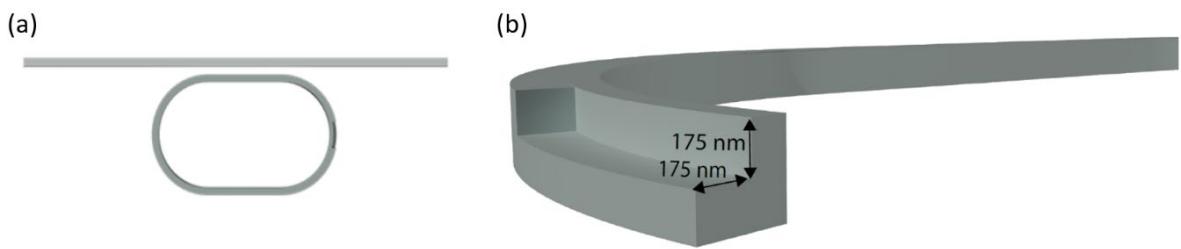

*Figure 3 a) Layout of ring resonator to couple with waveguide b) Structure of notch.*

that the radius of curved part, which is exactly a half circle, is 8.00 μm. By placing the notched structure such a way that its straight sides become parallel to waveguide that is 250 nm away from it, results mentioned above can be used to determine coupling length. 108.20 μm is found to be the length for which both polarizations are transferred to notched structure. The notch has width and height of both 175 nm as shown in Figure 3-b, and is used with the intend of providing interchange of energy between horizontal and vertical polarization modes. To implement different angles of half-wave plates, it is necessary to take the advantage of the fact that the rotation angles can be implemented by manipulating the notch size of the structure. To decide notch length, change of the quantity of horizontal and vertical polarizations at output port with respect to notch size is shown at Figure 4.

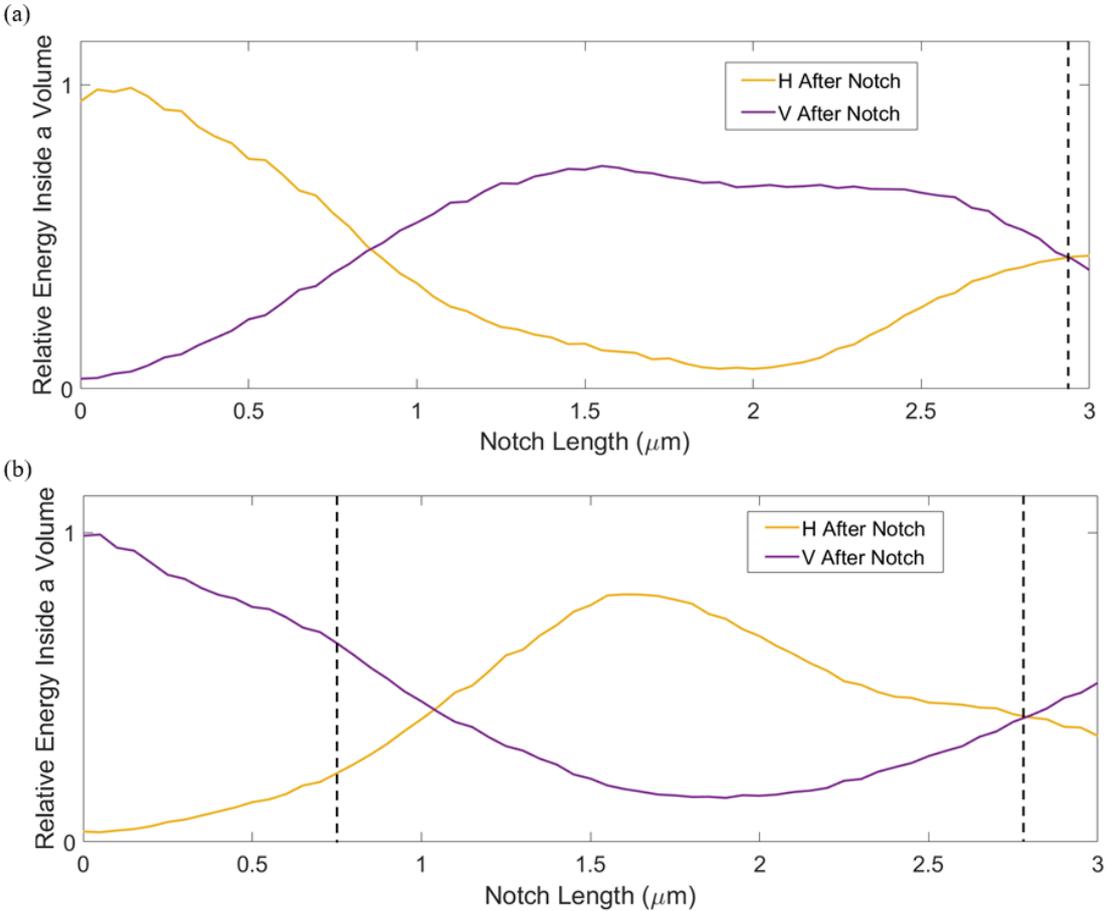

*Figure 4 Change of quantity of horizontally and vertically polarized light after the notch with respect to notch length when a) horizontally polarized b) vertically polarized light is initially sent to the structure. Appropriate lengths are marked by dashed lines.*

For HWP$_1$, by taking into consideration of only vertical polarization since horizontal polarization is not possible to reach HWP$_1$, the notch size is decided to be 0.75 μm to $\frac{1}{4}$ of output energy belong to horizontal polarization mode while rest stay in vertical polarization mode. On the other hand, for HWP$_2$ and HWP$_3$ input has both horizontal and vertical components and different notch lengths should be used for different input polarization. It can be achieved by first splitting two polarization components and then sending them to different notched ring resonators. Structure shown in Figure 5 is suitable for this purpose. From our previous coupling calculations, length of initial directional coupler (DC) is decided to be 70.72 μm. Notch length of ring resonator at upper arm (NRR$_1$) where horizontal polarization is processed should be 2.90 μm, while that of ring resonator at lower arm (NRR$_2$) should be 2.75 μm. For both case, output of notched ring resonators consists of equal energy of horizontal and vertical polarization states. All wave distributions for both polarizations and three notch lengths mentioned above are shown in Figure 6.

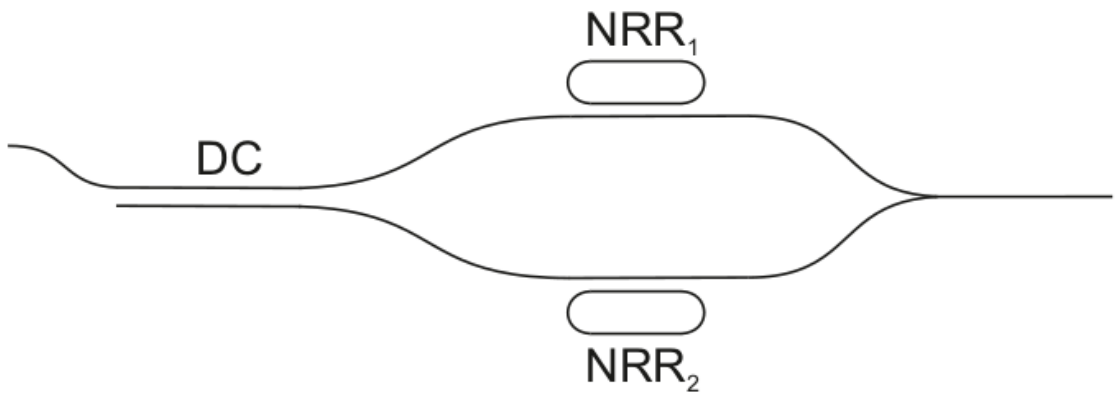

*Figure 5 Scheme of HWP$_2$ and HWP$_3$ with a directional coupler (DC) and two notched ring resonators (NRR$_1$ and NRR$_2$).*

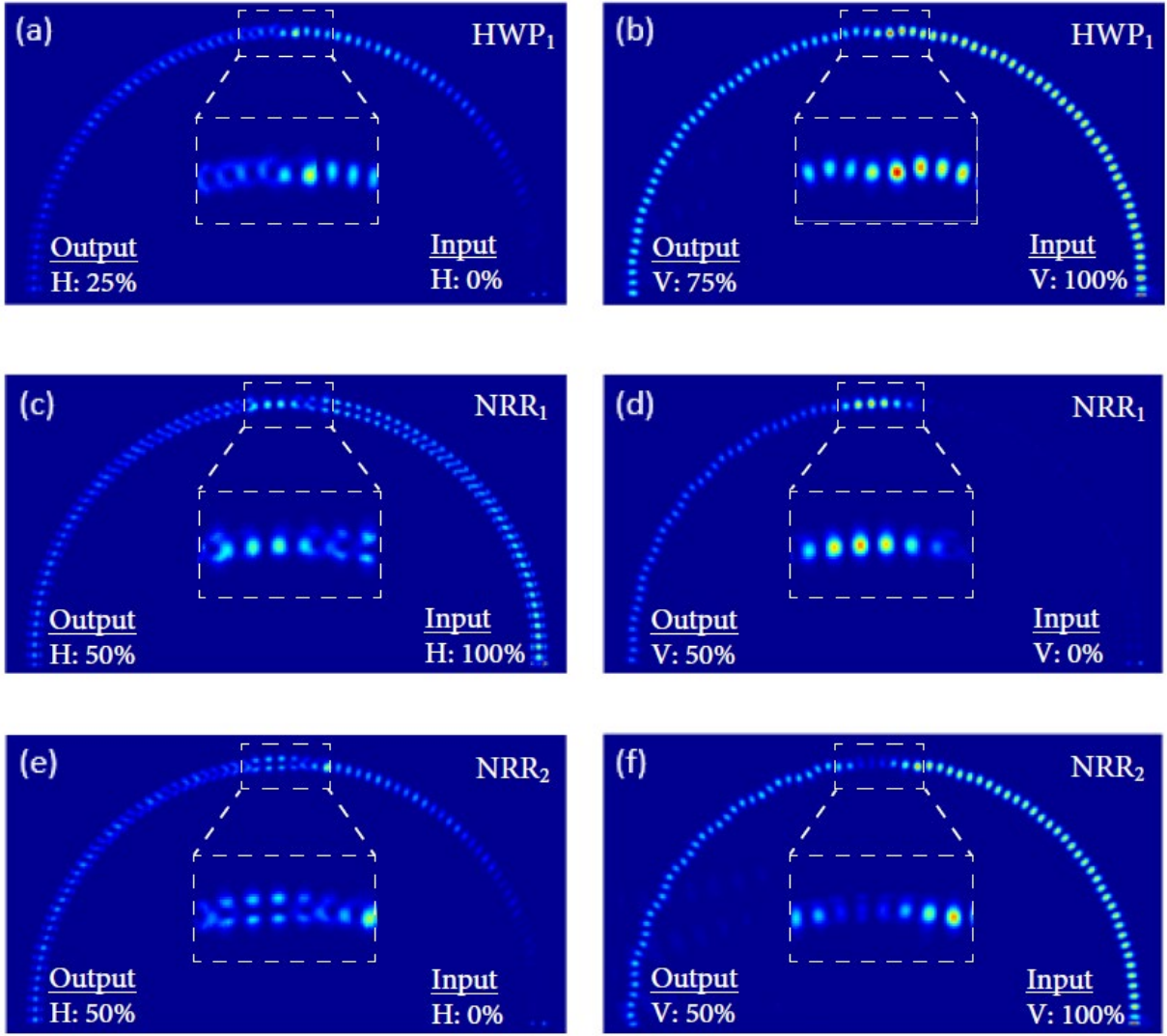

*Figure 6 Wave distributions for two polarizations inside notched ring resonators, including magnification of notch region and percentages of horizontal (H) and vertical (V) polarizations at input and output. a) Horizontal polarization b) vertical polarization when vertical polarization is sent to structure with 0.75 μm notch length. c) Horizontal polarization d) vertical polarization when horizontal polarization is sent to structure with 2.9 μm notch length. e) Horizontal polarization f) vertical polarization when vertical polarization is sent to structure with 2.75 μm notch length.*

Lastly, to ensure feasibility of fabricating such an integrated circuit, we perform simulations for slightly different dimensions. In Figure 7 response of 1 μm coupler to alteration of waveguide height, waveguide width and distance between two waveguides are shown, for dimension differences varying from -10 nm to 10 nm. In Figure 8, deviations from desired outcomes of notched ring resonators by change of dimensions of notch height, notch width, waveguide height and waveguide width by same amount as previous case are shown. Today's fabrication techniques enable us to produce waveguides with precision in such range [41], [42]. As can be seen in figures, there are tolerable deviations from desired operation.

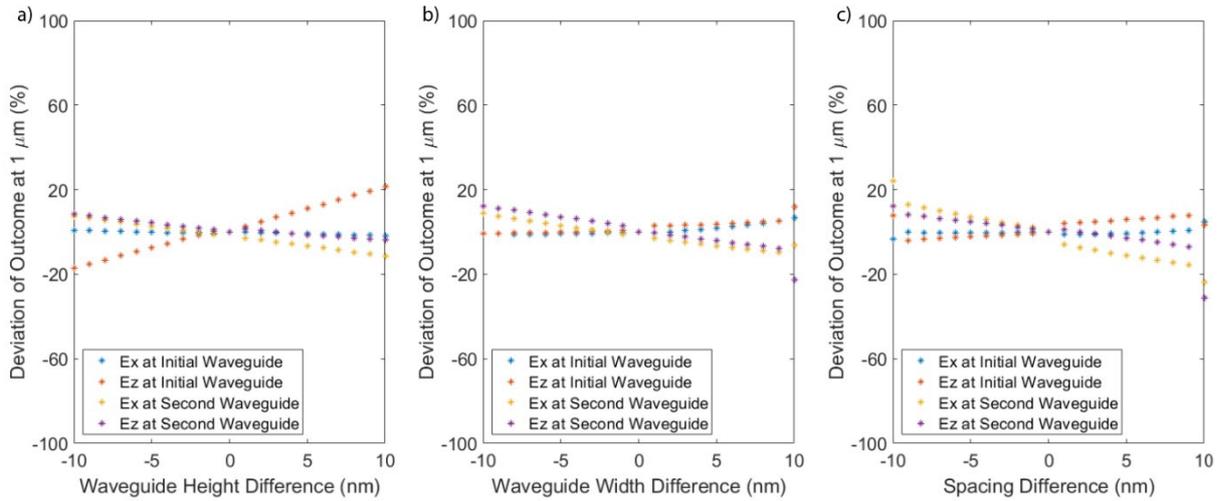

*Figure 7 Response of coupler to change of a) waveguide height b) waveguide width c) space between waveguides.*

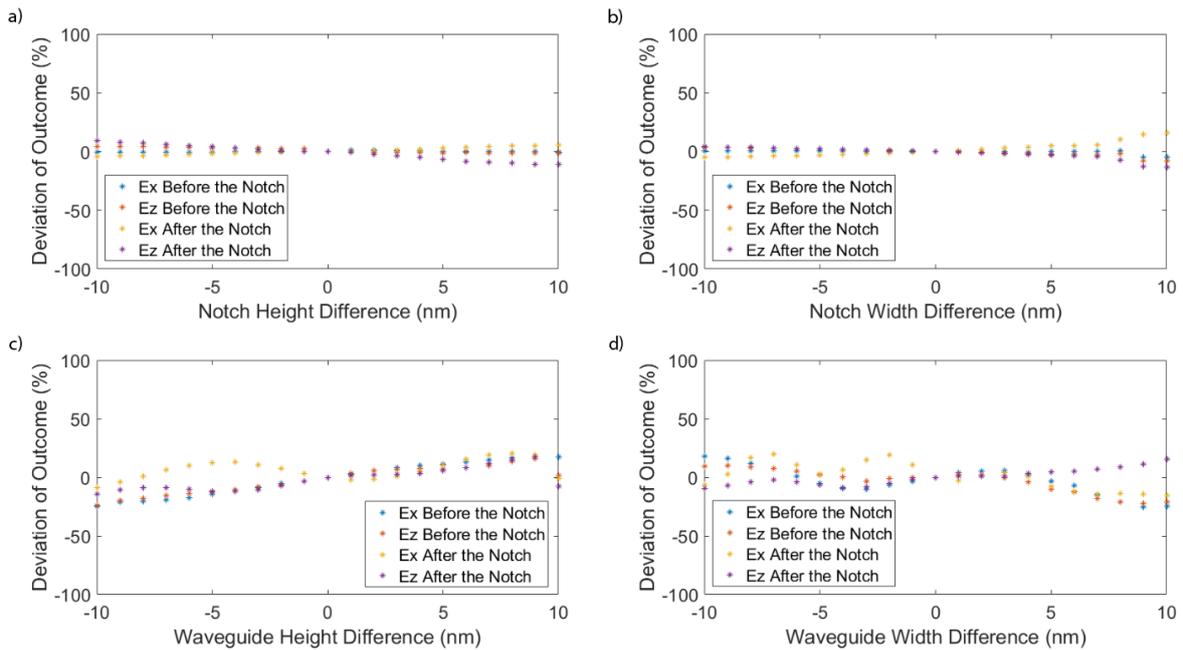

*Figure 8 Response of notched ring resonator to change of a) notch height b) notch width c) waveguide height d) waveguide width.*

## Conclusion

In conclusion, we proposed a programmable controlled phase gate to be used in photonic integrated circuits for quantum information processing. As a polarization rotator, we used a unique notched ring resonator that can be practically fabricated and is compatible with rectangular waveguide circuits. All of the elements proposed above can be feasibly fabricated thanks to the techniques as deep ultraviolet lithography or e-beam lithography [43]-[45]. Success probability of the gate is turns out to be $\frac{1}{48}$, which can be increased to $\frac{1}{12}$ by adding a feed-forward phase modulation [35]. Our compact integrated design will pave the way for large scale quantum information processing by providing opportunity of more versatile quantum operations.